\documentclass[letterpaper,twocolumn,english,aps,prb,floatfix,
showpacs, amsfonts, amssymb]{revtex4}
\usepackage[T1]{fontenc}
\usepackage[latin1]{inputenc}
\usepackage{amsmath}
\usepackage{graphicx}
\usepackage{amssymb}

\makeatletter

\providecommand{\LyX}{L\kern-.1667em\lower.25em\hbox{Y}\kern-.125emX\@}

\usepackage{subfigure}

\usepackage{amscd}
\usepackage{bm}

\usepackage{babel}
\makeatother
\begin{document}

\title{Phase diagrams and universality classes of random antiferromagnetic
spin ladders}

\author{J. A. Hoyos}

\email{joseabel@ifi.unicamp.br}

\affiliation{Instituto de Física Gleb Wataghin, Unicamp, Caixa Postal 6165, 13083-970.
Campinas, SP, Brazil}

\author{E.~Miranda}

\email{emiranda@ifi.unicamp.br}

\affiliation{Instituto de Física Gleb Wataghin, Unicamp, Caixa Postal 6165, 13083-970.
Campinas, SP, Brazil}

\date{\today{}}

\begin{abstract}
The random antiferromagnetic two-leg and zigzag spin-1/2 ladders are
investigated using the real space renormalization group scheme and
their complete phase diagrams are determined. We demonstrate that
the first system belongs to the same universality class of the dimerized
random spin-1/2 chain. The zigzag ladder, on the other hand, is in
a random singlet phase at weak frustration and disorder. Otherwise,
we give additional evidence that it belongs to the universality class
of the random antiferromagnetic and ferromagnetic quantum spin chains,
although the universal fixed point found in the latter system is never
realized. We find, however, a new universal fixed point at intermediate
disorder.
\end{abstract}


\maketitle

\section{Introduction}

One dimensional spin systems have been extensively studied over the
last several years and a fairly deep understanding of their possible
phases and corresponding physical behavior has emerged. Although actual
realizations are restricted to systems with a high degree of anisotropy,
one of the main motivations for these studies is the possibility of
cataloging their universality classes. This is specially tempting
in the case of disordered systems, since specially powerful methods
can be employed in one dimension, which are able to expose not only
average values but full distribution functions. One of the most studied
random one-dimensional spin systems is the random antiferromagnetic
(AF) spin-1/2 chain.\cite{mdh,fisher94} Making use of the real space
renormalization group (RSRG) method of Ma, Dasgupta and Hu,\cite{mdh}
it has been shown that, for any amount of uncorrelated disorder,\cite{dotyfisher}
the low-energy physics of this system is governed by an infinite randomness
fixed point.\cite{fisher94} The approach to this fixed point is characterized
by the formation at decreasing energy scales of random singlets between
widely separated spins. In this random singlet (RS) phase several
physical properties are universal and known.\cite{fisher94} For example,
the spin susceptibility $\chi \sim 1/T\log ^{2}T$, and the spin-spin
correlation function $C_{ij}=\left\langle \mathbf{S}_{i}\cdot \mathbf{S}_{j}\right\rangle $
is such that its mean value $\overline{C_{ij}}\sim \left(-1\right)^{i-j}/\left|i-j\right|^{2}$,
while the typical one $\left|C_{ij}\right|_{typ}\sim \exp (-\sqrt{\left|i-j\right|})$.

Other random systems have also been analyzed by these methods, among
which two are of special interest to us. One is the random dimerized
AF chain with different distributions for odd and even links.\cite{hyman1}
The presence of a gap in the clean version of this system provides
protection against the introduction of disorder. Therefore, for weak
disorder, the system retains a gap (or a pseudogap) and a spin susceptibility
which decreases to zero with decreasing temperature. For strong enough
disorder, however, the pseudogap is destroyed and the susceptibility
diverges as a power law. The power law exponent is non-universal and
varies continuously with the disorder strength, characterizing a Griffiths
phase. The other system of interest is the random chain with both
AF and ferromagnetic (FM) interactions.\cite{westerberg,hikihara}
It has been shown, using a generalization of the RSRG procedure, that
the low-energy behavior of this system is governed by the formation
of long clusters with large total spins. For weak disorder, this large
spin (LS) phase is universal. The magnetic susceptibility diverges
like a Curie law, and the specific heat vanishes like $T^{1/z}|\ln T|$,
with $1/z\approx 0.44$. However, for strong disorder, the LS phase
is no longer universal: although the magnetic susceptibility is still
Curie-like, the specific heat exponent $z$ varies continuously with
disorder.

More recently, following the discovery of spin ladder materials,\cite{spinladders}
attention has been drawn to coupled pairs of spin chains. The random
two-leg and zigzag ladders are the most studied of them.\cite{melin,yusuf,yusuf2}
The two-leg ladder is known to possess two phases:\cite{melin,yusuf}
a gapped phase, with vanishing spin susceptibility $\chi \left(T\right)$,
and a Griffiths phase, where $\chi \left(T\right)$ diverges as a
power law with continuously varying exponents as $T\to 0$. The zigzag
ladder, on the other hand, is topologically equivalent to single chains
with both nearest neighbor (nn) and next nearest neighbor (nnn) interactions.
Initial studies were confined to fairly small systems and concluded
in favor of the existence of only one phase of the Griffiths type.\cite{melin}
Subsequent investigations identified the presence of a small RS phase
for weak nnn interactions.\cite{yusuf2} Furthermore, the formation
of large effective spins, with FM and AF nn and vanishing nnn couplings
at the late stages of renormalization were taken as an indication
of a LS phase.\cite{yusuf2}

The main purpose of this work is to establish in detail the effective
low-energy equivalence between the random two-leg and zigzag ladders
with the three random spin chains mentioned above. We do so by means
of the RSRG method. In particular, we will show that at low energies
the random two-leg ladder is equivalent to the dimerized AF spin-1/2
chain. In the course of this analysis, we will construct the full
phase diagram of this model and confirm the existence of only two
possible phases. Moreover, we will show that there are indeed two
possible low-energy behaviors in the zigzag ladder. For small and
weakly disordered nnn couplings, it is equivalent to a random AF spin-1/2
chain with its RS phase. If the nnn interactions are stronger, however,
the system presents all the characteristic features of the LS phase
of a random chain with both AF and FM couplings. We thus confirm the
results of Ref.~\onlinecite{yusuf2} by showing the same scaling
of the effective cluster sizes and total spin values with energy as
seen in those systems. However, the zigzag ladders can never exhibit
the universal behavior observed in those systems at weak disorder.
This is demonstrated by an exhaustive investigation of the dependence
of the dynamical exponent $z$ on the shape and strength of the disorder
distribution. Nevertheless, we do find \emph{different} universal
regions where $z$ remains fixed while the disorder is varied. The
totality of our results thus serves to show that the universality
classes of random spin ladders can already be found in simpler systems
with nn interactions only.

The paper is organized as follows. In Sec.~\ref{zigzag} we introduce
the models we studied. The numerical results of the RSRG procedure
of both the conventional two-leg and zigzag ladders are presented
in Sec.~\ref{sec:Numerical-Results}. The unexpected behavior of
the dynamical exponent of the zigzag ladders is explained in Sec.~\ref{sec:t-model}.
We end with some final conclusions in Sec.~\ref{sec:Conclusions}.
Appendix~\ref{sec:calculationofz} discusses our method of calculation
of the dynamical exponent $z$ and a brief discussion of the case
of correlated disorder is given in Appendix~\ref{sec:Correlated-Disorder}.

\section{The Model\label{zigzag}}

We consider the following Hamiltonian, which describes an AF spin-1/2
chain with nearest neighbor (nn) and next nearest neighbor (nnn) interactions
\begin{equation}
H=\sum _{i=1}^{N-1}J_{i}\mathbf{S}_{i}\cdot \mathbf{S}_{i+1}+\sum _{i=1}^{N-2}K_{i}\mathbf{S}_{i}\cdot \mathbf{S}_{i+2},\label{2viz}\end{equation}
 where $\mathbf{S}_{i}$ is a spin-1/2 operator, $N$ is the total
number of spins, and $J_{i}>0$ e $K_{i}>0$ are the nn and nnn random
coupling constants, respectively. If $J_{i}=0$ for even $i$, this
is the two-leg ladder Hamiltonian; if $J_{i}$ is in general non-zero,
we have the zigzag ladder. The non-zero coupling constants $J_{i}$
and $K_{i}$ are in general independent random variables (see, however,
Appendix~\ref{sec:Correlated-Disorder}). We take them to be respectively
distributed in a power-law fashion (unless otherwise noted): \begin{subequations}\label{distinic}\begin{eqnarray}
P_{J}(J) & = & \frac{1-\alpha }{J_{0}}\left(\frac{J_{0}}{J}\right)^{\alpha }\; \, \, ,\; \, \, 0<J<J_{0},\label{distinic1}\\
P_{K}(K) & = & \frac{1-\alpha }{K_{0}}\left(\frac{K_{0}}{K}\right)^{\alpha }\; \, \, ,\; \, \, 0<K<K_{0}.\label{distinic2}
\end{eqnarray}
\end{subequations} The exponent $\alpha $ ($0\leq \alpha <1$) is
a measure of the disorder strength and the ratio of cutoffs $\Lambda =K_{0}/J_{0}$
gives the typical relative strength between nnn and nn interactions.

In order to study these systems we employed the RSRG method introduced
by Ma, Dasgupta and Hu.\cite{mdh} Its decimation steps consist in
isolating the strongest bond of the system ($\Omega $), keeping only
the lowest energy level of the bond, and renormalizing the remaining
interactions by perturbation theory. The new renormalized coupling
constants can be either ferromagnetic or antiferromagnetic. The details
of the procedure have been extensively discussed in the published
literature\cite{westerberg,melin,yusuf,yusuf2} and will be skipped
here. As the largest energy scale is lowered from its initial value
$\Omega _{0}=\mathrm{max}\left(J_{0},K_{0}\right)$, an effective
distribution of coupling constants is generated, which eventually
flows towards a fixed point distribution. The low energy behavior
of the system is governed by the remaining `active' non-decimated
spins at the scale of interest.\cite{mdh,fisher94,westerberg}

When there are initially only nn interactions, further neighbor interactions
are never generated.\cite{mdh,fisher94,westerberg} By contrast, in
our case, the range of effective couplings rapidly increases as the
RSRG is iterated (see Section~\ref{sec:t-model} for more details).
However, as the fixed point is approached, interactions beyond nearest
neighbors become extremely weak. If we then neglect interactions weaker
than a certain upper bound ($\Omega _{min}\approx 10^{-200}\Omega _{0}$)\cite{omegamin}
the effective range actually extends only as far as the nearest neighbors.
Thus, at the final stages of the RSRG, the ladders renormalize to
effective nearest neighbor spin chains.

\section{Numerical Results\label{sec:Numerical-Results}}

In this section we show the numerical results obtained from the iteration
of the RSRG for the two-leg and the zigzag ladders.

\subsection{Two-leg ladders}

\label{sub:Two-Leg-Ladders}

We first focus on the two-leg ladders ($J_{2i}=0$). In our simulations
we used chain lengths up to $N=200,000$.

\begin{figure}[h]
\begin{center}\includegraphics[  width=3in,
  keepaspectratio]{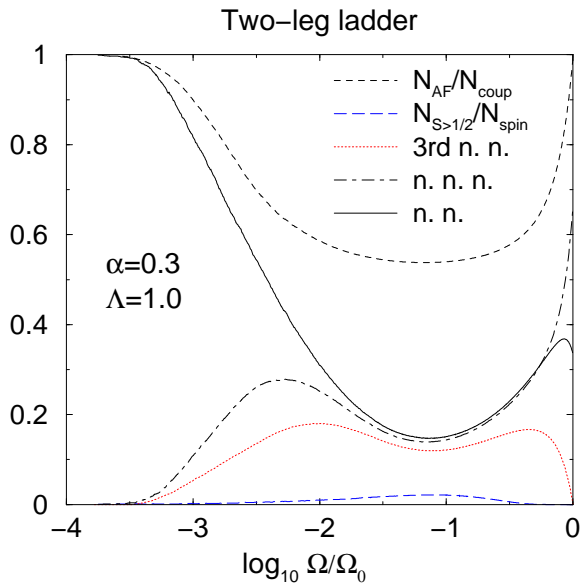}\end{center}

\caption{\label{fig1} The behavior of the fractions of nearest-neighbor,
next-nearest-neighbor, and third-nearest-neighbor coupling constants,
the fraction of spins greater than 1/2, and the fraction of antiferromagnetic
couplings as functions of the energy scale $\Omega $. The initial
number of spins is $N=200,000$, the disorder strength is $\alpha =0.3$,
and the ratio $\Lambda =1.0$. The RSRG is iterated until only nearest-neighbor
couplings are left. The data are averaged over 100 samples and the
relative error is less than $2\%$.}
\end{figure}

In Fig.~\ref{fig1}, we show the behavior of the fractions of nn,
nnn, and 3rd nn bonds as functions of the energy scale $\Omega $.
In addition, this Figure also shows the fraction of `active' spins
greater than 1/2 and the fraction of antiferromagnetic coupling constants
as functions of $\Omega $. The first thing to note is that the only
significant couplings at the lowest energy scales ($\alt 10^{-3.5}\, \Omega _{0}$)
are nn couplings. Furthermore, these remaining couplings are all antiferromagnetic.
This has been verified for all values of $\alpha $ and $\Lambda $.

\begin{figure}[h]
\begin{center}\includegraphics[  width=3in,
  keepaspectratio]{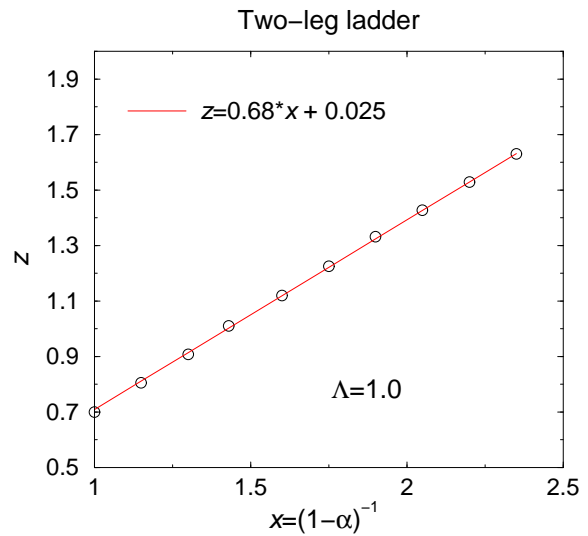}\end{center}

\caption{\label{fig2}Variation of the dynamical exponent $z$ with the disorder
strength $\alpha $ for $\Lambda =1.0$. $N=200,000$ is the initial
number of spins in the ladder and the RSRG is iterated until there
are only nearest-neighbor bonds left. The data are average over 5
samples and the relative error is about $3\%$, comparable to the symbol
size.}
\end{figure}

Another important feature of the approach to the fixed point is the
difference between the distributions of the odd bonds and the even
ones. We have checked that, at the lowest energy scales, all the even
bonds vanish while the odd ones are distributed according to\begin{equation}
P_{Jodd}(J)\sim J^{-1+1/z},\label{deltadist}\end{equation}
 where $z$ is the dynamical exponent.\cite{fisher95,melin} Thus,
the system renormalizes to a collection of \emph{free effective random
dimers}. The average magnetic susceptibility is then given by\cite{fisher94,hyman1}\[
\chi \sim T^{1-z}.\]
 If $z<1$, the density of states is suppressed at low energies and
the system has a `soft gap' (or a pseudo-gap). This is a remnant of
the Haldane-type gap\cite{haldaneconj} of the clean two-leg ladder.\cite{scalapino}
We therefore call it a disordered Haldane phase. Otherwise, the system
shows no such suppression and is in a gapless (Griffiths) phase.\cite{melin}
We have determined the value of $z$, for different disorder strengths
(see the Appendix~\ref{sec:calculationofz} for details on the method
of computation of $z$). As shown in Fig.~\ref{fig2}, for $\Lambda =1$,
a transition between these two phases occurs when $\alpha \approx 0.3$.
In both phases the system is strongly dimerized. Within the length
scale of the size of the effective dimers, the correlations decay
as a power law. We believe this decay is similar to the case of random
Heisenberg chains ($\sim 1/r^{2}$), since there are equal contributions
coming from even and odd bond decimations. At larger length scales,
the correlations are strongly (exponentially) suppressed. This overall
behavior is apparent in Fig.~5 of Ref.~\onlinecite{yusuf}. The
crossover between the two regimes is governed by the effective dimer
sizes, which are primarily determined by $\Lambda $ not $\alpha $.

\begin{figure}[h]
\begin{center}\includegraphics[  width=3in,
  keepaspectratio]{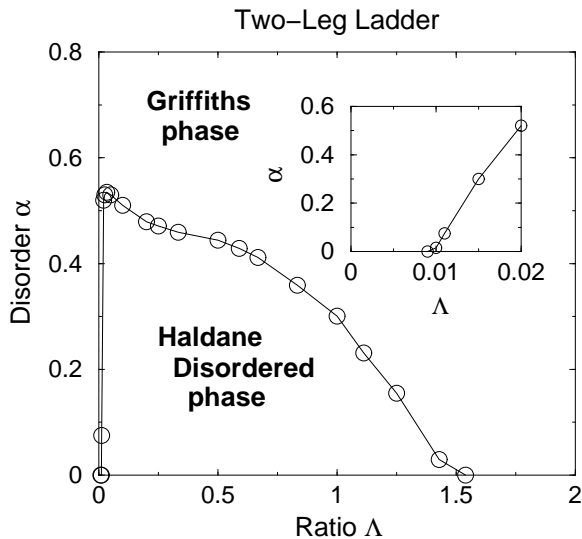}\end{center}

\caption{\label{fig3}Phase diagram of the random two-leg ladder. The disordered
Haldane phase is characterized by a `soft gap' (or pseudo-gap),
whereas in the Griffiths phase, the disorder completely destroys this
remnant of the Haldane gap. In both phases the system is
dimerized. The data error is about the size of the symbols.}
\end{figure}

Calculating the dynamical exponent $z$ for various values of $\Lambda $
and $\alpha $, one can construct the phase diagram of the two-leg
ladder (Fig.~\ref{fig3}). In the transition line, the dynamical
exponent equals one and the low-energy density of states and the magnetic
susceptibility as $T\rightarrow 0$ are both constant.

{}From the inset of Fig.~\ref{fig3}, we note that for $\Lambda \alt 10^{-2}$
the system is always in the Griffiths phase. We can understand this
in the following manner. In the limit of $\Lambda \rightarrow 0$,
the system reduces to a collection of disconnected dimers (the `rungs'
of the ladder), whose couplings are distributed according to Eq.~(\ref{distinic1}).
Therefore, $z=\left(1-\alpha \right)^{-1}\geq 1$ and the system is
always in a Griffiths phase. For small $\Lambda $, this behavior
is preserved. For $10^{-2}<\Lambda <1.53$, the Haldane-type gap of
the clean system gives rise to a `soft gap' upon the introduction
of weak disorder. For large disorder, the Griffiths phase re-emerges,
with a diverging non-universal magnetic susceptibility. For $\Lambda >1.53$,
only the Griffiths phase exists. This behavior smoothly connects with
the $\Lambda \to \infty $ limit of two disconnected random Heisenberg
chains, which is governed by the infinite randomness fixed point.\cite{fisher94}
Formally, this limit corresponds to $z\to \infty $. We note that
for any value of $\Lambda $, there are only two types of phases.\cite{melin,yusuf}

We stress the high degree of similarity between the disordered two-leg
ladder and the dimerized random antiferromagnetic chain,\cite{hyman1}
which also shows analogous phases to the ones discussed here. Indeed,
in both cases the RSRG flow leads to a fixed point with nn antiferromagnetic
interactions only but with different distributions for even and odd
bonds. Thus, the two systems clearly belong to the same universality
class.

\begin{figure}[h]
\begin{center}\includegraphics[  width=3in,
  keepaspectratio]{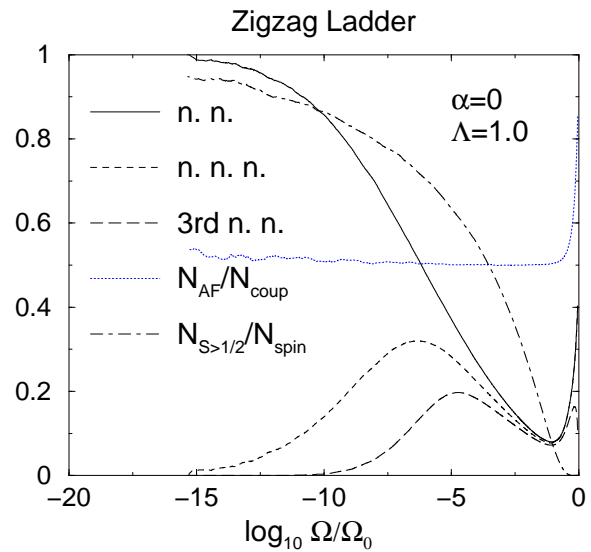}\end{center}

\caption{\label{fig4} The fraction of nearest-neighbor, next-nearest-neighbor,
third-nearest-neighbor, the fraction of spins greater than 1/2, and
the fraction of AF couplings for zigzag ladders as functions of the
energy scale $\Omega $. The initial number of spins is $N=640,000$,
the disorder strength is $\alpha =0$ and the ratio $\Lambda =1.0$.
The renormalization group is iterated until all the bonds in the
system are between nearest-neighbors only. The data are averaged over
5 samples.  For $\Omega /\Omega _{0}\geq 10^{-10}$ the data relative
error is less than $2\%$, otherwise, it is less than $8\%$.}
\end{figure}

\subsection{Zigzag ladders}

\label{sub:Zigzag-Ladders}

We now consider the zigzag ladders, with coupling constants distributed
according to Eq.~(\ref{distinic}). We studied ladders with initial
lengths of $N=640,000$.

Yusuf and Yang\cite{yusuf2} have calculated the ratio of the average
nn interactions to that of further neighbor interactions in zigzag
ladders with correlated disorder. They have shown that this ratio
increases tremendously as the energy scale is lowered. Indeed, if
we plot the fraction of nn, nnn, and 3rd nn interactions as functions
of $\Omega $, for $\alpha =0$ and $\Lambda =1$ (see Fig.~\ref{fig4}),
we can see that only nn bonds survive at the lowest energies ($\approx 10^{-15}\, \Omega _{0}$).
This is similar to the case of two-leg ladders, as shown in Section
\ref{sub:Two-Leg-Ladders}, and was verified in our simulations, irrespective
of the values of $\alpha $ and $\Lambda $. However, in contrast
to the two-leg ladder case, the nn bonds are \emph{not} all antiferromagnetic.
The asymptotic number of ferromagnetic and antiferromagnetic bonds
is about the same, as shown also in Fig.~\ref{fig4}. Finally, as
also pointed out in Ref.~\onlinecite{yusuf2}, there is a rapid proliferation
of spins greater than $1/2$. This is typical of the so-called Large
Spin (LS) phase found in disordered Heisenberg chains with both FM
and AF interactions.\cite{westerberg} As we will see, this similarity
is not fortuitous.

\begin{figure}[h]
\begin{center}\includegraphics[  width=3in,
  keepaspectratio]{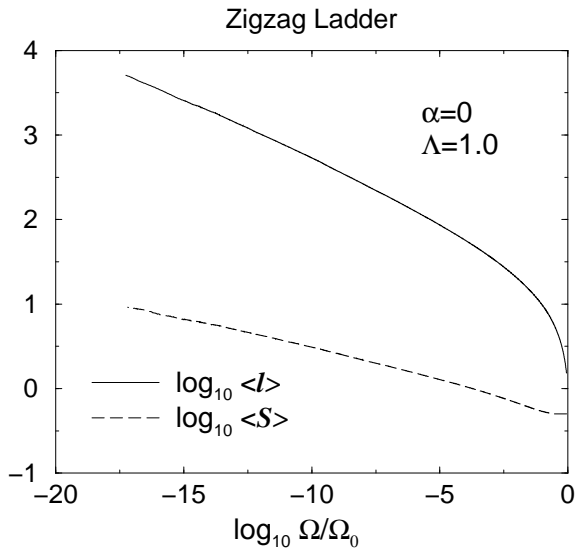}\end{center}

\caption{\label{fig5} The average cluster size $<l>$ and the average spin
size $<S>$ as functions of the energy scale $\Omega $. From a direct
fit we get $<l>\sim \Omega ^{-0.14}$ and $<S>\sim \Omega ^{-0.069}$.
The initial length of the system is $N=640,000$, $\alpha =0$ and
$\Lambda =1.0$. The renormalization group is iterated until only
nearest-neighbor couplings are left. The data were averaged over 5
samples and the sample to sample variation is less than $5\%$.}
\end{figure}

One of the distinguishing features of the disordered chains with both
FM and AF interactions is the relation between the average spin size
and the average cluster length. The average cluster size is related
to the energy scale through the dynamical exponent $z$ by $<l>\sim \rho ^{-1}\sim \Omega ^{-1/z}$
and the average spin size grows with the lowering energy scale according
to $<S>\sim \Omega ^{-\kappa }$. Both behaviors are observed in the
zigzag ladders as shown in Fig.~\ref{fig5}. As first pointed out
by Westerberg \emph{et al.},\cite{westerberg} the two exponents are
related by $1/z=2\kappa $ in random chains with both FM and AF interactions.
This follows from the fact that the main decimation process is not
singlet formation but rather the formation of large spins from the
random addition and subtraction of spin pairs. As a consequence, the
cluster growth is characterized by a random walk in `spin size space'.
Again, we find this relation also holds for the zigzag ladders ($\kappa \approx 0.069$
and $1/z\approx 0.14$ for the chain of Fig.~\ref{fig5}). This is
evidence that the zigzag ladders belong to the same universality class
of the disordered chains with FM and AF interactions. 

Our simulations also show that the fixed point distributions of FM
and AF link excitation gaps are the same, with a characteristic power
law dependence\cite{deltadef}\begin{equation}
P_{AF}\left(\Delta \right)=P_{F}\left(\Delta \right)=P\left(\Delta \right)\sim \Delta ^{-1+1/z},\label{eq:zzgaps}\end{equation}
 where $z$ is the same dynamical exponent obtained from the relation
between $<l>$ and $\Omega $ (see below for an explanation).

An interesting feature of the zigzag ladders is the following. In
general, the total fixed point distribution of link excitation gaps
is a linear combination of AF and FM contributions, namely,\begin{eqnarray*}
P(\Delta ) & = & \frac{1}{\Omega }\left[\frac{x}{z_{AF}}\left(\frac{\Omega }{\Delta _{AF}}\right)^{1-1/z_{AF}}\right.\\
 &  & \left.+\frac{1-x}{z_{FM}}\left(\frac{\Omega }{\Delta _{FM}}\right)^{1-1/z_{FM}}\right],
\end{eqnarray*}
where the first and second terms come from the AF and FM link distributions,
respectively, and $x$ is the fraction of AF couplings. In our simulations,
the fraction of AF bonds is $x\approx 0.53$ in the LS phase. The
relation between length scale and the energy scale is given by\cite{fisher94}\begin{equation}
\frac{d\rho }{d\Omega }=P(\Delta =\Omega )\rho .\label{eq:rate}\end{equation}
The coefficient in front of $P\left(\Delta =\Omega \right)$ is taken
to be 1 because the main decimation process replaces 2 spins by 1.
Solving Eq.~(\ref{eq:rate}) we get \[
\rho \sim \Omega ^{1/z}\]
with \begin{equation}
1/z=x/z_{AF}+(1-x)/z_{FM}.\label{eq:zrel}\end{equation}
This general relation seems to have been unnoticed in previous studies.
In particular, we note that weakly disordered chains with both AF
and FM couplings have $1/z_{FM}\approx 0.56$, $1/z_{AF}\approx 0.30$,
and $x\approx 0.63$,\cite{westerberg} so that $1/z\approx 0.40$
which is reasonably close to the value of $0.44\pm 0.02$ found directly
in the simulations of Ref.~\onlinecite{westerberg}. In contrast,
in the zigzag ladders, we have found $z_{AF}=z_{FM}$, irrespective
of the value of $x$ ($x\approx 0.53$ in the LS phase). As a result,
$z_{AF}=z_{FM}=z$.

\begin{figure}[h]
\begin{center}\includegraphics[  width=3in,
  keepaspectratio]{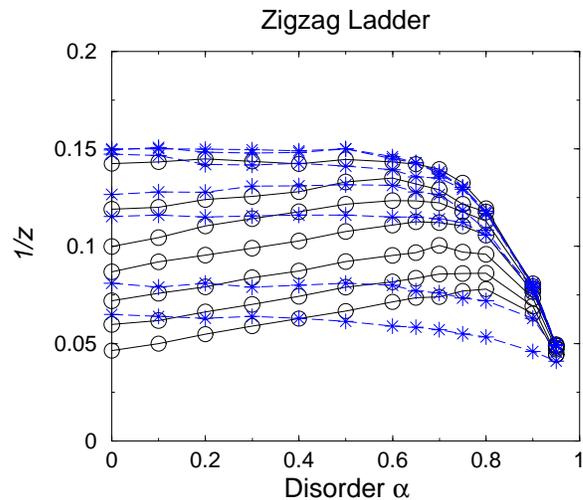}\end{center}

\caption{\label{fig6}Variation of the dynamical exponent $z$ with the disorder
parameter $\alpha $ for different values of $\Lambda $. From top to
bottom, the solid lines with circles correspond to $\Lambda =1.0$,
0.6, 0.4, 0.2, 0.1, 0.05, and 0.0025, whereas the dashed lines with
stars are for $\Lambda =0.8^{-1}$, $0.6^{-1}$, $0.4^{-1}$, $0.2^{-1}$,
$0.1^{-1}$, $10^{3}$, and $10^{6}$. $N=640,000$ is the initial number
of spins in the ladder and the renormalization group is iterated until
the bonds in the system are nearest-neighbor only. The data relative
error is less than $10\%$, about the size of the symbols.}
\end{figure}

Now, we would like to explore the disorder dependence of the dynamical
exponent $z$. Fig.~\ref{fig6} illustrates the behavior of $z$
as a function of the disorder parameter $\alpha $ for various values
of $\Lambda $. Note that the smallest value of $z$ is $1/0.15$.
Westerberg \emph{et al.} have found that for a range of weak disorder
strengths, random chains with both FM and AF couplings are characterized
by a universal value of $z_{U}=1/0.44$.\cite{westerberg} Strongly
singular disorder distributions, however, show non-universal disorder-dependent
values of $z$. As can be seen from Fig.~\ref{fig6}, for initial
power-law distributions of couplings, zigzag ladders always have $z>z_{U}$
and are never in the basin of attraction of the universal behavior.

For $\alpha \leq 0.6$ the behavior of $1/z$ is linear in $\alpha $
\[
1/z=a\alpha +b.\]
 Note that, for $\Lambda \geq 1.0$, $a=0$ and for $1.0<\Lambda <1/0.4$
the system flows to a \emph{new universal fixed point} where $1/z=0.15\pm 0.02$.
This is similar to the random AF and FM chain\cite{westerberg} but
the value of the dynamical exponents are different. We note from the
general trend of the curves in Fig.~\ref{fig6} that the origin of
this universal behavior seems to be the non-monotonic behavior of
$z$ as one goes from $\Lambda \alt 1$ to $\Lambda \agt 1$ (see
also Fig.~\ref{fig7} below). This behavior, on the other hand, can
be understood from simple physical arguments as shown below in Section~\ref{sec:t-model}.

For $\Lambda \leq 0.4$ all the lines have the same slope $a=0.039\pm 0.001$
and the intercept $b$ varies in a logarithm manner with $\Lambda $,
i.e., $b=b_{1}+b_{2}\ln \Lambda $, where $b_{1}=0.12\pm 0.01$ and
$b_{2}=0.020\pm 0.001$. Thus, there is a phase transition at the
value of $\alpha $ where the dynamical exponent diverges, $\alpha =\alpha _{c}\left(\Lambda \right)=-\left(b_{2}\ln \Lambda +b_{1}\right)/a$.
For $\alpha \leq \alpha _{c}\left(\Lambda \right)$, the system is
governed by an infinite randomness fixed point where the magnitude
of the spins does not grow and the FM couplings vanish. Indeed, the
presence of such a RS phase was observed by Yusuf and Yang\cite{yusuf2}
by direct calculation of $z$. Hence, we conclude that the system
presents two phases. In the greater part of phase diagram it is in
the LS phase, with a new universal fixed point in the region $1.0<\Lambda <1/0.4$
and $\alpha <0.6$. In addition, there is a tiny region where the
system is in a RS phase.

For $\alpha >0.6$, all the curves converge to the point $\alpha =0.95$
and $z=1/0.05$. Note in this connection that $\alpha =0.95$ corresponds
to an initial distribution which is the same as the fixed point one
if $z=1/0.05$, cf. Eqs.~(\ref{distinic1}) and (\ref{eq:zzgaps}).
This is a reflection of the inability of the decimation procedure,
which is dominated by first order perturbation theory steps, to generate
distributions which are more singular than the initial one.

\begin{figure}[h]
\begin{center}\includegraphics[  width=3in,
  keepaspectratio]{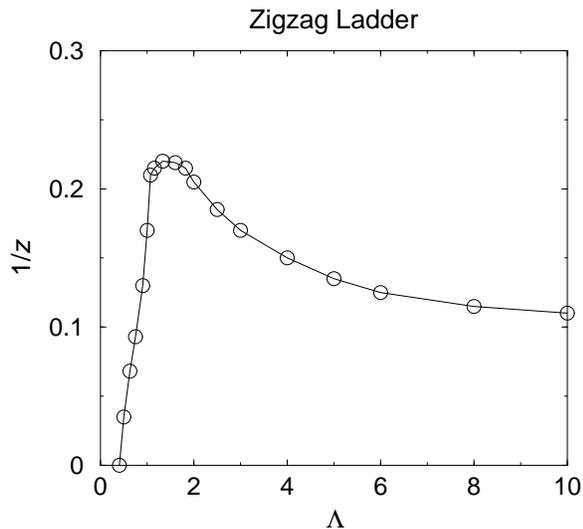}\end{center}

\caption{\label{fig7}Variation of the dynamical exponent $z$ with the ratio
$\Lambda $ for initial `box distributions with a gap' ($\delta =0.05$,
see text for details). $N=200,000$ is the initial number of spins in
the ladder and the renormalization group is iterated until the bonds
in the system are nearest-neighbor only. The data relative error is
less than $5\%$, about the symbol size.}
\end{figure}

One might think that the universal behavior of disordered chains with
AF and FM interactions where $z=z_{U}$ could be realized in zigzag
ladders with disorder distributions which are not as singular as Eq.~(\ref{distinic}).
This is not the case, however, as exemplified by the case of a `box
distribution with a gap', where the initial distributions are uniform
and have support in $J_{0}-\delta <J<J_{0}$ and $K_{0}-\delta <K<K_{0}$
(we take $\max \{J_{0},K_{0}\}=1$). We have explored several values
of $\delta $ and found $z$ to be always greater than $z_{U}$. This
is seen, for example, in the extreme case of $\delta =0.05$ in Fig.~\ref{fig7},
where we plot $1/z$ as a function of $\Lambda $. Note that there
is a phase transition at $\Lambda =0.4$. To the left of this point,
the system is in a RS spin-1/2 phase and the dynamical
exponent diverges $z\sim -\ln \Omega $. For $\Lambda >0.4$, the
system is in the LS phase. The smallest value of the dynamical exponent
is $z=1/0.22$ (still greater than the universal $z_{U}$ found by
Westerberg \textit{et. al}) and for $\Lambda \agt 10$, $z$ saturates
at $z=1/0.11$. We conclude that the LS phase of the zigzag ladders
is in the same universality class of random FM and AF chains with
strong disorder but it does not reach the universal region found in
those chains when the disorder is weak. It should be remembered that
the clean system is gapless if $\Lambda \alt 0.24$ but spontaneously
dimerized and gapful if $\Lambda \agt 0.24$.\cite{halddimer} The
topological nature of the dimerized state, however, makes it unstable
with respect to any finite amount of disorder and the random system
is always gapless.\cite{yang}

\begin{figure}[h]
\begin{center}\includegraphics[  width=3in]{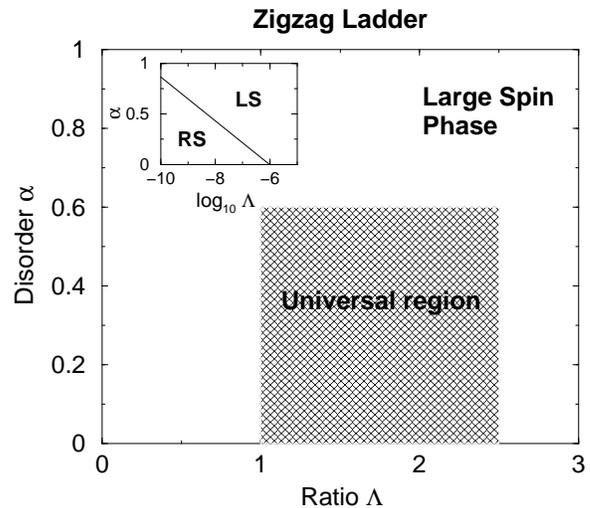}\end{center}

\caption{Phase diagram of the random zigzag ladder. The Large Spin phase
dominates most of the parameter space. There is a universal region
(the shaded area) for $1\alt \Lambda \alt 2.5$ and $0\alt \alpha \alt
0.6$ characterized by $z\approx 1/0.15$.  There is also a tiny region
for very small values of $\Lambda $ (see the inset), where the system
is in a Random Singlet phase.\label{fig8}}
\end{figure}

Based on the behavior of the dynamical exponent $z$, we can determine
the phase diagram of the random zigzag ladder, as shown the
Fig.~\ref{fig8}.  The Large Spin phase is dominant in most of the
($\alpha $,$\Lambda $) parameter space.  In this phase, the low energy
physics of the system is governed by a fixed point where the mean size
of the spin clusters grows when the energy scale is lowered, and they
are weakly coupled. The distribution of link excitations gaps is not
universal except in the shaded region. This universal region in the
Large Spin phase is characterized by a dynamical exponent $z\approx
1/0.15$. The thermodynamic properties are well known:\cite{westerberg}
the specific heat $C\sim T^{1/z}|\ln T|$ and there is a Curie-like
magnetic susceptibility $\chi \sim T^{-1}$.  The average spin-spin
correlation function decays in a logarithmic manner.\cite{hikihara} In
addition to this phase, there is a tiny region where the system is in
a RS phase (where $z\rightarrow \infty $, see the inset of the
Fig.~\ref{fig8}). Here, the low energy physics is governed by a
universal infinite randomness fixed point.  The specific heat $C\sim
-1/\ln ^{3}T$, the magnetic susceptibility $\chi \sim 1/\left(T\ln
^{2}T\right)$, the mean correlation function $C_{m}(r)\sim r^{-2}$,
and the typical correlation function $C_{t}(r)\sim \exp \left\{
-r^{-1/2}\right\} $.\cite{fisher94} This phase was previously
identified in Ref.~\onlinecite{yusuf2}, although its precise location
differs somewhat from our results. We attribute this difference to
finite-size effects and the different methods used for the
characterization of RS behavior.

\section{Anomalous disorder dependence of $z$ in the zigzag ladders}

\label{sec:t-model}

An intriguing aspect of the data shown in Fig.~\ref{fig6} is the
fact that $z$ decreases as the initial disorder strength $\alpha $
is increased, for $\Lambda <1$ and $\alpha <0.6$ (solid lines of
Figure~\ref{fig6}). This behavior is unexpected since the weaker
the initial disorder is, the stronger is the final effective one.
As we will show, this anomalous behavior can be simply understood
by analyzing the limiting behavior as $\Lambda \to 0.$

\begin{figure}[h]
\begin{center}\includegraphics[  width=3in,
  keepaspectratio]{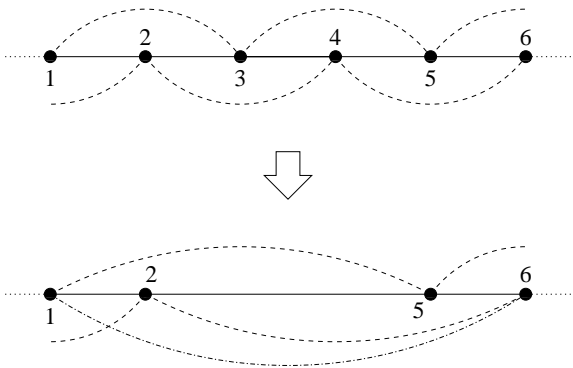}\end{center}

\caption{\label{fig9}Schematic decimations at the earliest stages of the
RG.}
\end{figure}

Let us start by writing down the effective couplings generated by
the decimation procedure in its earliest stages when all the spins
are spin-1/2 and the couplings are AF (see Fig.~\ref{fig9}). Let
us assume that the coupling between spins $S_{3}$ and $S_{4}$ is
the largest one in the system, i.e., $\Omega =J_{34}$. After the
decimation step, both spins are removed from the system because they
are effectively frozen in a singlet state. The remaining spins are
coupled with renormalized interactions as follows. The nn couplings
are\begin{eqnarray*}
\tilde{J}_{12} & = & J_{12}-\frac{K_{13}\left(J_{23}-K_{24}\right)}{2\Omega },\\
\tilde{J}_{25} & = & \frac{\left(J_{23}-K_{24}\right)\left(J_{45}-K_{35}\right)}{2\Omega },\\
\tilde{J}_{56} & = & J_{56}-\frac{K_{46}\left(J_{45}-K_{35}\right)}{2\Omega },
\end{eqnarray*}
 the nnn ones are \begin{eqnarray*}
\tilde{K}_{15} & = & \frac{K_{13}\left(J_{45}-K_{35}\right)}{2\Omega },\\
\tilde{K}_{26} & = & \frac{K_{46}\left(J_{23}-K_{24}\right)}{2\Omega },
\end{eqnarray*}
 and a third-nearest-neighbor coupling between spins $S_{1}$ and
$S_{6}$ is generated \[
\tilde{L}_{16}=\frac{K_{13}K_{46}}{2\Omega }.\]

First, we analyze the limit $\Lambda \ll 1$. In this case, it is
easy to see that the renormalized nn, nnn, and 3rd nn couplings are
of order $\mathcal{O}(\Lambda ^{0})$, $\mathcal{O}(\Lambda ^{1})$,
and $\mathcal{O}(\Lambda ^{2})$, respectively. Thus, if we neglect
the 3rd nn coupling, the original form of the Hamiltonian is recovered,
and the nn couplings remain stronger than the nnn ones. No FM coupling
is generated and as the energy scale is lowered the nnn couplings
vanish faster than the nn ones. The system flows to a random AF spin-1/2
chain in a RS phase. This scheme breaks down when a nn FM coupling
appears leading to frustration. Among the three nn renormalized couplings,
$\tilde{J}_{25}$ is the most likely to be FM with probability given
by \begin{eqnarray*}
P\left(\tilde{J}_{25}<0\right) & = & 2P\left(J_{23}<K_{24}\right)P\left(J_{45}>K_{35}\right)\\
 & = & 2\left\{ \int _{0}^{\Lambda }P_{J}\left(J\right)\left[\int _{J}^{\Lambda }P_{K}\left(K\right)dK\right]dJ\right\} \times \\
 &  & \left\{ 1-\int _{0}^{\Lambda }P_{J}\left(J\right)\left[\int _{J}^{\Lambda }P_{K}\left(K\right)dK\right]dJ\right\} \\
 & = & \Lambda ^{1-\alpha }\left(1-\frac{1}{2}\Lambda ^{1-\alpha }\right).
\end{eqnarray*}
 As expected, this probability is greater, the greater $\Lambda $
is, since $\Lambda $ governs the strength of frustration. But note
also that it increases with $\alpha $. As more FM couplings are produced,
the RG flow tends to deviate from the RS phase. As a result, the dynamical
exponent will tend to be smaller. This explains the anomalous behavior
we have found.

In the opposite limit ($\Lambda \gg 1$), this picture no longer holds.
Suppose that $\Omega =K_{35}$. In this case, the renormalized coupling
between spins $S_{2}$ and $S_{4}$ is now a nn one \[
\tilde{J}_{24}=K_{24}+\frac{J_{23}(J_{45}-J_{34})}{2\Omega },\]
 while the nnn coupling between $S_{1}$ and $S_{4}$ is \[
\tilde{K}_{14}=\frac{K_{13}(J_{45}-J_{34})}{2\Omega }.\]
 It is clear that in this case the nn coupling tends to increase and
the nnn one has a better chance of becoming a FM coupling in the earliest
stages of the RG flow. This will rapidly enhance the frustration of
the system and drive it away from the RS phase. In this limit, we
expect the RS phase only at $\Lambda ^{-1}=0$.

The competing tendencies at small and large $\Lambda $ lead to a
minimum value of $z$ (Figs.~\ref{fig6} and \ref{fig7}). The presence
of this minimum partially explains the universal region where $z$
is approximately constant at $\approx 6.7$ (Fig.~\ref{fig6}).

A similar type of reasoning can be used to analyze the case of correlated
disorder considered in Ref.~\onlinecite{yusuf2}. This is discussed
in Appendix~\ref{sec:Correlated-Disorder}.

\section{Conclusions\label{sec:Conclusions}}

We have studied random AF two-leg and zigzag spin-1/2 ladders using
the real space renormalization group method of Ma, Dasgupta and Hu.
We have determined the complete phase diagrams of these two general
models in great detail by calculating the dynamical exponent $z$
and the range of low-energy effective interactions. The two-leg ladders
show a gapped disordered Haldane region with a random dimer nature
and a gapless Griffiths phase, but no random singlet phase. The zigzag
ladder, on the other hand, can be either in a random singlet phase
or a large spin phase.

One of our main findings is that throughout their phase diagrams these
two models lead to effective low-energy models with nearest neighbor
interactions only. This simplification had been noticed before in
connection with the zigzag ladders.\cite{yusuf2} Our calculations
confirm those results and generalize them to the two-leg ladders.
Furthermore, by analyzing the structure of odd and even links in the
case of the two-leg ladders and the scaling with energy of the large
clusters of spins in the case of the zigzag ladders, we have been
able to show the low-energy equivalence of these systems to the random
dimerized AF spin chain and the random chain with AF and FM interactions,
respectively. The low-energy equivalence between random systems with
further neighbor yet short-ranged interactions to systems with nearest
neighbor interactions only is likely to hold in general. However,
there remains the possibility that one-dimensional models with longer
ranged interactions are not decimated down to nearest neighbor spin
chains in the process of the RG.\cite{yusuf2} It would be interesting
to determine the critical range beyond which such simplification does
not occur.

While the zigzag ladder could be related to the random chain with
both AF and FM chains, we have shown that their phase diagrams do
not completely overlap. Indeed, the latter system has a universal
fixed point at weak disorder (with $z\approx 2.27$)\cite{westerberg}
which is inaccessible to the former. The zigzag ladder, on the other
hand, has a region of parameter space with a new universal behavior:
for initial distributions less singular than $P_{J}\left(x\right)\sim P_{K}\left(x\right)\sim x^{-\alpha _{c}}$,
with $\alpha _{c}\approx 0.6$, and for $1.0\lesssim \Lambda \alt 1/0.4$,
the dynamical exponent is always $z\approx 6.7$. This critical disorder
strength $\alpha _{c}$ is comparable to the one found in the random
AF and FM chain,\cite{westerberg} although we see no other similarities
between these two regimes. We have been able to find a rough explanation
for this universal behavior by ascribing it to competing tendencies
which create a shallow valley where $z$ is approximately constant.
It is possible that the universal regime of Ref.~\onlinecite{westerberg}
has a similar origin.

\begin{acknowledgments}
We would like to acknowledge financial support from FAPESP through
grants 03/00777-3 (JAH) and 01/00719-8 (EM) and CNPq 301222/97-5 (EM). 
\end{acknowledgments}
\appendix

\section{The calculation of the dynamical exponent $z$}

\label{sec:calculationofz}

\begin{widetext}

\begin{figure*}[h]
\begin{center}\includegraphics[  width=3in]{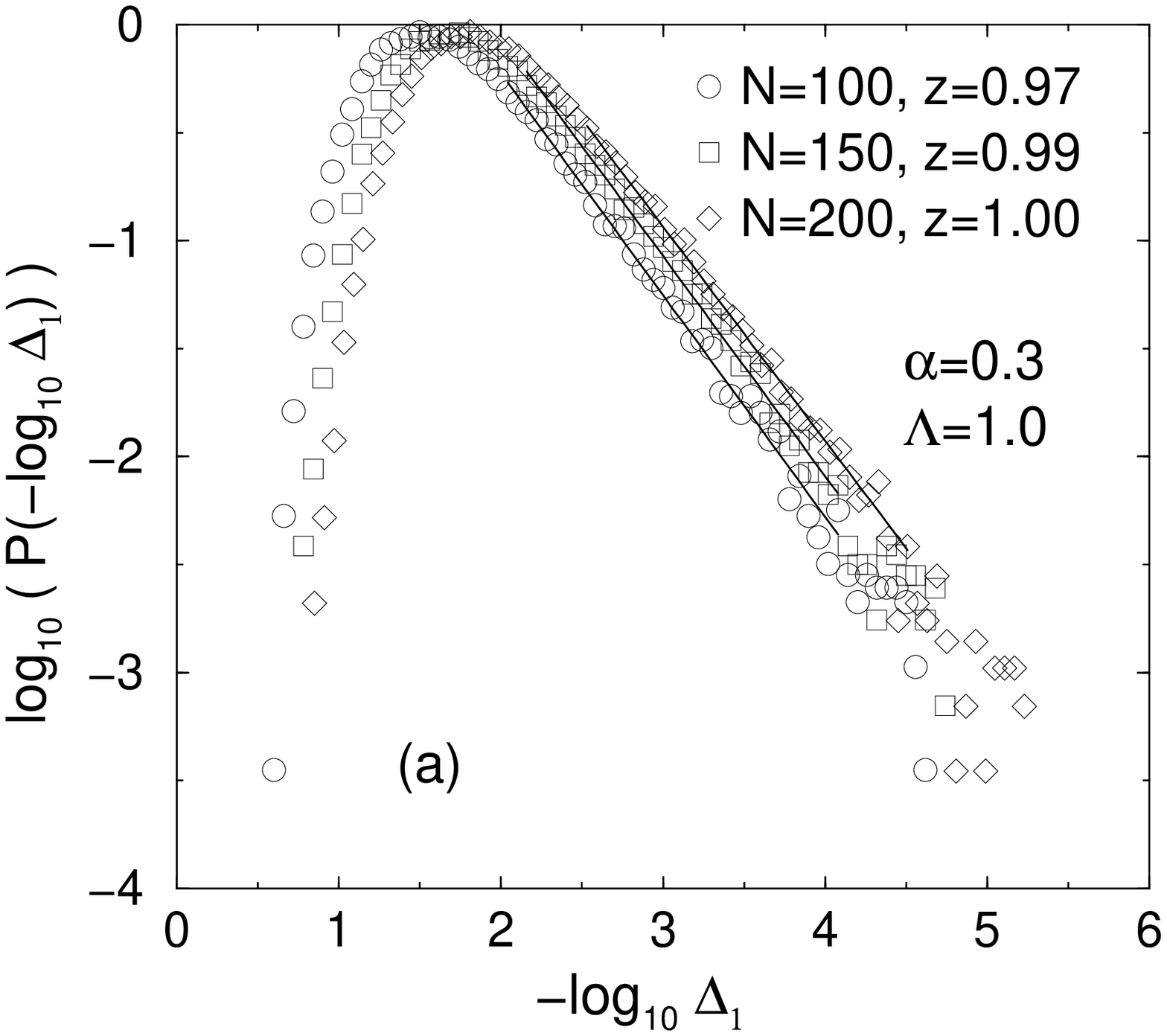} \includegraphics[  width=3in]{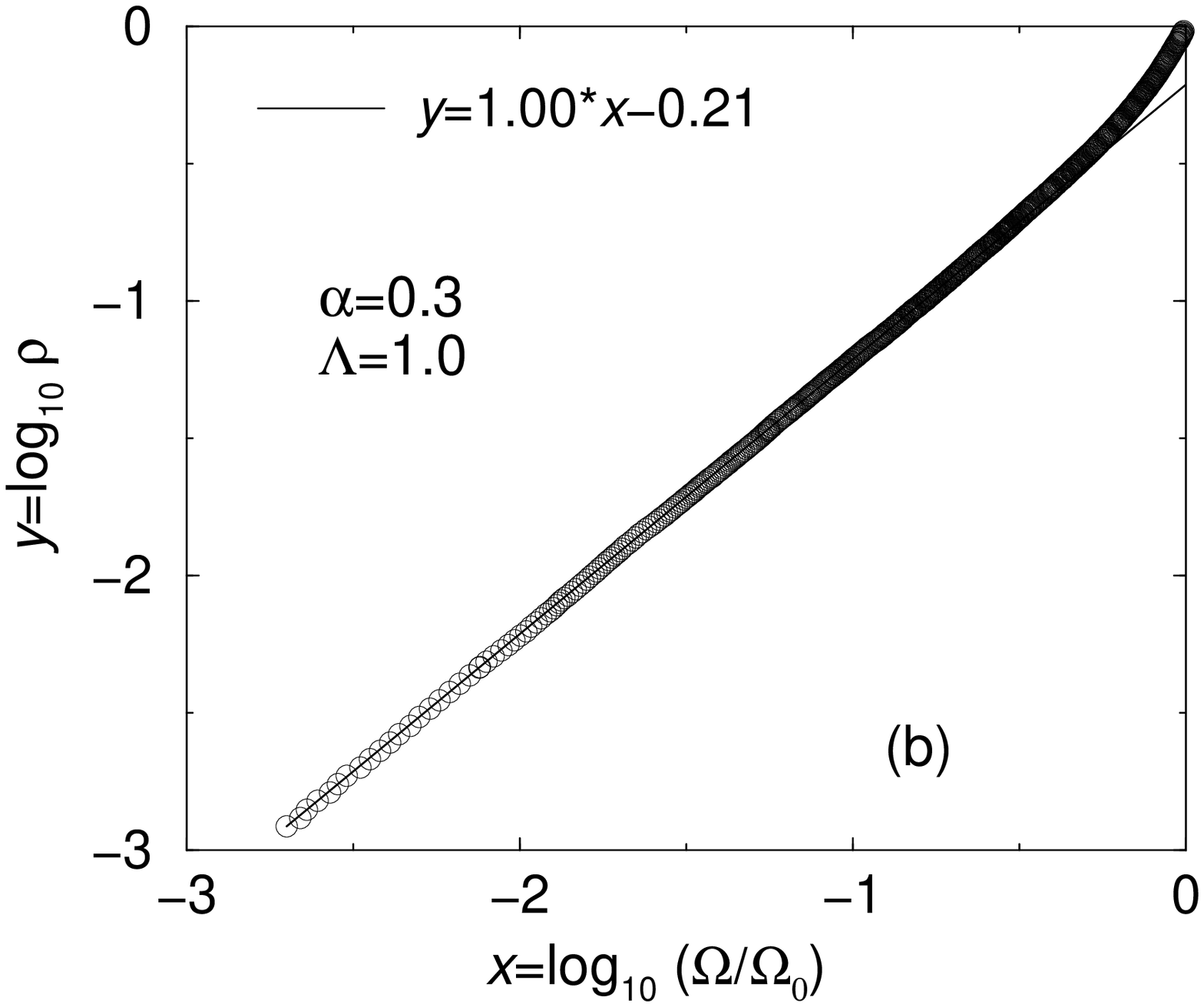}\end{center}

\caption{\label{fig10}The calculation of the dynamical exponent $z$. (a)
The distribution of first gaps $\Delta _{1}$ using 50,000 realizations
for $N$=100, 150, and 200, is fitted to $\log _{10}P(-\log _{10}\Delta
_{1})=\mathrm{const.}+\left(1/z\right)\log \Delta _{1}$, and (b) $z$
is directly calculated from $\log _{10}\rho \sim
\mathrm{const.}+\left(1/z\right)\log _{10}\Omega $, where we used
$N$=200,000. The data are from the decimation of a \emph{single}
chain. We verified that the variation with respect to other
realizations of disorder is less than $1\%$.}
\end{figure*}

\end{widetext}

In this appendix, we would like to comment on our method of calculation
of the dynamical exponent $z$. A common procedure, frequently used
in the literature, is to decimate the system until one pair of spins
is left. One can then calculate the excitation gap of this last dimer
(the first gap $\Delta _{1}$). By repeating this procedure for different
realizations of disorder one can obtain the distribution of first
gaps. The dynamical exponent is obtained by fitting the behavior at
small $\Delta _{1}$ to the power law in Eq.~(\ref{deltadist}).\cite{melin,saguia}
This is shown in Fig.~\ref{fig10}(a). Another option is to calculate
the density of `active' spins $\rho $ at the scale $\Omega $. The
dynamical exponent $z$ is then obtained from its definition, through
the relation between the length scale $l\sim \rho ^{-1}$ and the
energy scale $\Omega $ \begin{equation}
\rho \sim \Omega ^{1/z}.\label{expdin}\end{equation}

This second method is exemplified in Fig.~\ref{fig10}(b), and the
agreement is excellent. While the methods are equivalent it should
be pointed out that the second method is computationally much faster,
as \emph{a single} realization of a large chain is sufficient for
the determination of $z$. In our calculations, we have chosen to
use the second method. This was actually crucial in the case of the
zigzag ladders, where the asymptotic behavior can only be obtained
with very large system sizes,\cite{melin} rendering the first method
very inefficient. This need for large sizes is due to the presence
of long-range correlations in these systems\cite{hikihara} as opposed
to the exponential dependence of the two-leg ladders.\cite{yusuf}

\section{Zigzag ladders with correlated disorder}

\label{sec:Correlated-Disorder}

Yusuf and Yang have also considered the case of zigzag ladders with
correlations between the random nn and nnn couplings.\cite{yusuf2}
They have considered nn couplings distributed according to Eq.~(\ref{distinic1})
and nnn couplings given by \[
K_{i}=\Lambda \frac{J_{i}J_{i+1}}{\Omega _{0}}.\]
Their analysis identified a RS phase for $\Lambda <0.5$, and a LS
phase for $0.5<\Lambda <1.0$.

We can qualitatively analyze this case along the same lines as in
Sec.~\ref{sec:t-model}. Suppose $\Omega =J_{34}$ in Fig.~\ref{fig9}.
Using the above definition of $K_{i}$ in the earliest stages of the
RG flow (when $\Omega \approx \Omega _{0}$), we have \[
\tilde{J}_{25}=(1-\Lambda )^{2}\frac{J_{23}J_{45}}{2\Omega },\]
 \[
\tilde{K}_{15}=\Lambda ^{\prime }\frac{J_{12}\tilde{J}_{25}}{\Omega },\]
 \[
\tilde{K}_{26}=\Lambda ^{\prime }\frac{\tilde{J}_{25}J_{56}}{\Omega },\]
 where $\Lambda ^{\prime }=\Lambda /(1-\Lambda )$. Neglecting the
3rd nn coupling and the renormalization of $J_{12}$ and $J_{56}$,
the new Hamiltonian has the same form as the original one, except
for the fact that the anisotropy parameter for $\tilde{K}_{15}$ and
$\tilde{K}_{26}$ has been renormalized. In the first RG decimations,
no FM couplings can arise. However, in a second run there is a finite
probability for FM couplings to appear if $\Lambda ^{\prime }>1$.
This happens if $\Lambda >0.5$. The presence of FM couplings introduces
frustration which is the mechanism that drives the system away from
the RS phase towards the LS phase.


\begin{thebibliography}{10}
\bibitem{omegamin}We take $\Omega _{min}$ to be well below the lowest energy scale
down to which we decimate ($\approx 10^{-40}\Omega _{\textrm{0 }}$),
so that the neglected couplings are utterly irrelevant. We have checked
that the results do not change if we use other values of $\Omega _{min}$
consistent with this constraint.
\bibitem{deltadef}We follow the convention of Ref.~\onlinecite{westerberg} and define
$\Delta $ as the energy difference between the ground and first excited
multiplets of a given pair of coupled spins.
\end{thebibliography}
\end{document}